\pdfoutput=1
\documentclass[10pt,twocolumn]{article}
\usepackage[pdftex,dvips]{graphicx}
\usepackage{times}

\newcommand{\uvec}{\mbox{\boldmath $U$\unboldmath}}

\def\be{\begin{eqnarray}}
\def\ee{\end{eqnarray}}
\def\benl{\begin{eqnarray*}}
\def\eenl{\end{eqnarray*}}
\def\dt{ \Delta t}
\def\dx{ \Delta x}
\def\dy{ \Delta y}
\def\dz{ \Delta z}
\newcommand{\nwc}{\newcommand}
\nwc{\bm}{\boldmath}
\nwc{\m}{\mbox}
\nwc{\ubm}{\unboldmath}
\nwc{\bmU}{\m{\bm$U$\ubm}}
\nwc{\bmX}{\m{\bm$X$\ubm}}
\nwc{\bmu}{\m{\bm$u$\ubm}}
\nwc{\bmx}{\m{\bm$x$\ubm}}
\nwc{\bmz}{\m{\bm$z$\ubm}}
\nwc{\bmv}{\m{\bm$v$\ubm}}
\nwc{\bmw}{\m{\bm$w$\ubm}}
\nwc{\bmW}{\m{\bm$W$\ubm}}
\nwc{\bmn}{\m{\bm$n$\ubm}}
\nwc{\bmG}{\m{\bm$G$\ubm}}
\nwc{\bmF}{\m{\bm$F$\ubm}}
\nwc{\bmI}{\m{\bm$I$\ubm}}
\nwc{\bmN}{\m{\bm$N$\ubm}}
\nwc{\bmP}{\m{\bm$P$\ubm}}
\nwc{\bmcalP}{\m{\bm $\cal P$\ubm}}
\nwc{\bmV}{\m{\bm$V$\ubm}}
\nwc{\bmS}{\m{\bm$S$\ubm}}

\usepackage[dvips]{graphicx}

\begin{document}

\title{\LARGE\bf The Numerical Simulation of Ship Waves \\
using Cartesian Grid Methods}

\author{{\Large Mark Sussman$^1$ and Douglas G.\ Dommermuth$^2$} \\ \\
$^1$Department of Mathematics, Florida State University, \\
Tallahassee, FL 32306 USA, sussman@zeno.math.fsu.edu \\ \\
$^2$Naval Hydrodynamics Division, Science Applications International 
Corporation, \\
10260 Campus Point Drive, MS 34, San Diego, CA  92121 USA, \\
douglas.g.dommermuth@saic.com}
\date{}
\maketitle

\begin{abstract}
Two different cartesian-grid methods are used to simulate the flow 
around the DDG 5415. The first technique uses a ``coupled level-set 
and volume-of-fluid'' (CLS) technique to model the free-surface 
interface.  The no-flux boundary condition on the hull is imposed 
using a finite-volume technique.  The second technique uses a 
level-set technique (LS) to model the free-surface interface.   A 
body-force technique is used to impose the hull boundary condition. 
The predictions of both numerical techniques are compared to 
whisker-probe measurements of the DDG 5415.  The level-set technique 
is also used to investigate the breakup of a two-dimensional spray 
sheet.
\end{abstract}

\section{Introduction}

At moderate to high speed, the turbulent flow along the hull of a 
ship and behind the stern is characterized by complex physical 
processes which involve breaking waves, air entrainment, free-surface 
turbulence, and the formation of spray.  Traditional numerical 
approaches to these problems, which use boundary-fitted grids, are 
difficult and time-consuming to implement.  Also, as waves steepen, 
boundary-fitted grids will break down unless ad hoc treatments are 
implemented to prevent the waves from getting too steep.  At the very 
least, a bridge is required between potential-flow methods, which 
model limited physics, and more complex boundary-fitted grid methods, 
which incorporate more physics, albeit with great effort and with 
limitations on the wave steepness. Cartesian-grid methods are a 
natural choice because they allow more complex physics than 
potential-flow methods, and, unlike boundary-fitted methods, 
cartesian-grid methods require minimal effort with no limitation on 
the wave steepness.   Although cartesian-grid methods (CGM) are 
presently incapable of resolving the hull boundary-layer, CGM can 
model wave breaking, free-surface turbulence, air entrainment, 
spray-sheet formation, and complex interactions between the ship hull 
and the free surface, such as transom-stern flows and tumblehome 
bows.   The cartesian-grid methods that are described in this paper 
use the panelized geometry that is used by potential-flow methods to 
automatically construct a representation of the hull.   The hull 
representation is then immersed inside a cartesian grid that used to 
track the interface.  No additional gridding beyond what is already 
used by potential-flow methods is required.   We note that another 
variation of this approach is to use cartesian-grid methods to track 
the free-surface interface and body-fitted grids to model the ship 
hull.

For the calculation of ship waves, VOF and level-set methods have 
certain advantages and disadvantages.  VOF uses the volume fraction 
($F$) to track the interface.   $F=0$ corresponds to gas and $F=1$ 
corresponds to liquid.   For intermediate values, between zero and 
one, there exists an interface between the gas and the liquid.   The 
interface between the gas and the liquid is sharp for a pure VOF 
method.  Level-set methods use a level-set function ($\phi$) to model 
the gas-liquid interface.  By definition, $\phi<0$ denotes gas, 
$\phi>0$ denotes liquid, and $\phi=0$ is the interface. For 
conventional level-set schemes, the interface between the gas and the 
liquid is given a finite thickness\cite{SusSmeOsh94}, which is unlike 
conventional VOF schemes\cite{BraKotZem92}.

In the case of free-surface flows, where the density ratio between 
air and water is almost three orders of magnitude, the finite 
thickness of the interface that characterizes level-set methods has 
two advantages over VOF.  First, the finite thickness tends to smooth 
jumps in the tangential component of the velocity on the interface. 
Second, the finite  thickness tends to facilitate using multigrid 
methods to solve various types of elliptic equations that involve the 
density.

The advection algorithm that is used for VOF conserves mass if the 
flow field is solenoidal.   The level-set advection equation tends to 
accumulate numerical errors.   For the level-set method, the 
level-set function must be periodically reinitialized to maintain a 
proper thickness for the interface, otherwise the interface would 
become either too thick or too thin.   The reinitialization process 
is a significant source of errors in the level-set method.   Based on 
accuracy considerations, the calculation of gravity-driven flows 
tends to favor VOF over level-set methods.

The interface is reconstructed from the volume fractions in VOF. 
During the reconstruction process, the interface normals and 
curvature are calculated.   Typically, the calculation of the 
interface normal and curvature are less accurate for VOF than for 
level-set methods.  The interface normals and curvature are 
calculated directly in level-set methods in terms of gradients of the 
level-set function.   As a result, the calculation of the normals and 
curvature are less costly for level-set methods relative to VOF. 
The calculation of surface tension effects, which are a function of 
the curvature of the interface, tends to favor level-set methods over 
VOF due to considerations of accuracy and efficiency.

On highly-stretched, multidimensional grids, VOF methods are less 
prone to aliasing errors than level-set methods.  Level-set methods 
incur errors as the interface rotates through highly resolved regions 
into regions that are not resolved well.   This type of aliasing 
error occurs in cartesian-grid methods when the mesh along one 
coordinate axis is more finely resolved than along another coordinate 
axis.

By definition, the level-set and volume-of-fluid function both allow 
mixing of gas and liquid.   This feature of level-set and 
volume-of-fluid methods may be desirable for modeling gas entrainment 
such as the air that is entrained by a breaking wave.   During the 
reinitialization process, level-set methods and ``coupled level set 
and volume-of-fluid methods'' (CLS) use a signed distance function to 
update the level-set function and the thickness of the interface. 
Naturally, the distance function could be used to model the intensity 
of turbulence and amount of gas entrainment as a function of the 
distance to the interface.

Dommermuth, et al., (1998) used a stratified flow formulation to 
simulate breaking bow waves on the DDG 5415 at a Froude number 
Fr=0.41.  Their numerical results compared well to whisker-probe 
measurements in the bow region \cite{DomInnLutNovSchTal98}. However, 
Dommermuth, et al., (1998) identified two issues that required 
further study.  First, their stratified flow formulation allowed the 
free-surface interface to become too diffuse.  Second, the 
contact-line treatment didnot allow the free surface to rise and fall 
cleanly along the side of the hull.  The two new numerical approaches 
that are discussed in this paper are attempts to remedy these 
problems.

Both numerical approaches use a signed distance function to represent 
the hull.   The distance of a point to the hull is negative inside 
the hull and positive outside the hull.   The finite-volume approach 
uses the signed distance to calculate the area and volume fractions 
for computational cells cut by the hull, whereas the body-force 
technique uses the signed distance to prescribe a smooth forcing 
term.  The coupled interface-tracking algorithm (CLS) uses level-set 
to calculate the normals (and curvature if needed) to the 
free-surface interface that are used in VOF.   The advection portion 
of the algorithm is performed by VOF \cite{SusPuc99}.  The level-set 
interface-tracking algorithm uses a new isosurface scheme to 
calculate the zero level-set.   Then the minimal distance between the 
cartesian points and the zero level-set is calculated in a narrow 
band.  The minimal distance is made positive in the water and 
negative in the air.  This signed distance to the free surface is 
used to reinitialize the thickness of the interface.

The two numerical approaches are used to simulate the flow around the 
DDG 5415.   The CLS technique is still under development, so only 
preliminary results are presented.  The level-set technique includes 
upgrades to the numerical technique that is described in 
\cite{DomInnLutNovSchTal98}. Those upgrades include a new body-force 
formulation that is mollified, a new reinitialization procedure, and 
a new finite-volume treatment of the convective terms.  The original 
numerical procedure is not mollified and does not use 
reinitialization.  In addition, the original central-difference 
formulation of the convective terms is not as robust as the new 
treatment using a flux integral formulation. We first review the 
governing equations and then we discuss the numerical approaches. 
Finally, we present some preliminary numerical results which 
illustrate various features of the numerical algorithms.  The 
application of level-set methods to the breakup of spray sheets is 
also illustrated.

\section{\label{sec:equations}Field Equations}

As in Dommermuth, et al., (1998), consider turbulent flow at the 
interface between air and water \cite{DomInnLutNovSchTal98}.  Let 
$u_i$ denote the three-dimensional velocity field as a function of 
space ($x_i$) and time ($t$).  For an incompressible flow, the 
conservation of mass gives

\begin{eqnarray}
\label{mass}
\frac{\partial u_i}{\partial x_i} = 0 \;\; .
\end{eqnarray}

\noindent  $u_i$ and $x_i$ are normalized by $U_o$ and $L_o$, which 
are the characteristic velocity and length scales of the body, 
respectively.  On the surface of the moving body ($S_b$), the fluid 
particles move with the body:

\begin{eqnarray}
u_i = U_i \;\; ,
\end{eqnarray}

\noindent where $U_i$ is the velocity of the body.

Let $V_\ell$ and $V_g$ respectively denote the liquid (water) and gas 
(air) volumes.  Following a procedure that is similar to 
\cite{OshSet88, SusAlmBelColHowWel99}, we let $\phi$ denote a 
level-set function.  By definition, $\phi=-1$ for ${\bf x} \in V_g$ 
and $\phi=1$ for  ${\bf x} \in V_\ell$. The fluid interface 
corresponds to $\phi=0$.

The convection of $\phi$ is expressed as follows:

\begin{eqnarray}
\label{level}
\frac{d \phi}{d t} = \frac{\partial Q}{\partial x_j} \;\; ,
\end{eqnarray}

\noindent where $d/dt=\partial/\partial t + u_i \partial/\partial 
x_i$ is a substantial derivative.  $Q$ is a sub-grid-scale flux which 
can model the entrainment of gas into the liquid.  Details are 
provided in \cite{DomInnLutNovSchTal98}.

Let $\rho_\ell$ and $\mu_\ell$ respectively denote the density and 
dynamic viscosity of water.  Similarly, $\rho_g$ and $\mu_g$ are the 
corresponding properties of air.  The flow in the water and air is 
governed by the Navier-Stokes equations:

\begin{eqnarray}
\label{navi}
\frac{d u_i}{d t} & = & F_i
-\frac{1}{\rho} \frac{\partial P}{\partial x_i}
+\frac{1}{\rho R_e} \frac{\partial}{\partial x_j} \left( 2 \mu S_{ij} \right)
\nonumber \\
& & -\frac{1}{F_r^2} \delta_{i3}
     +\frac{1}{\rho W_e} T_i
     +\frac{\partial  \tau_{ij}}{\partial x_j} \;\; ,
\end{eqnarray}

\noindent where $R_e=\rho_\ell U_o L_o/\mu_\ell$ is the Reynolds 
number, $F_r^2 = U_o^2/(g L_o)$ is the Froude number, and 
$W_e=\rho_\ell U_o^2 L_o/\sigma$ is the Weber number. $g$ is the 
acceleration of gravity, and $\sigma$ is the surface tension. $F_i$ 
is a body force that is used to impose boundary conditions on the 
surface of the body.  $P$ is the pressure.  $T_i$ accounts for 
surface-tension effects.  $\delta_{ij}$ is the Kronecker delta 
symbol.  As described in \cite{DomInnLutNovSchTal98}, $\tau_{ij}$ is 
the subgrid-scale stress tensor. $S_{ij}$ is the deformation tensor:

\begin{eqnarray}
S_{ij} & = & \frac{1}{2} \left(
\frac{\partial u_i}{\partial x_j}
+\frac{\partial u_j}{\partial x_i} \right) \;\; .
\end{eqnarray}

\noindent $\rho$ and $\mu$ are respectively the dimensionless variable
densities and viscosities:

\begin{eqnarray}
\rho(\phi) & = & \lambda + (1 - \lambda) {\rm H} (\phi) \nonumber \\
\mu(\phi) & = & \eta + (1 - \eta ) {\rm H} (\phi) \;\; ,
\end{eqnarray}

\noindent where $\lambda = \rho_g/\rho_\ell$ and $\eta = 
\mu_g/\mu_\ell$ are the density and viscosity ratios between air and 
water. For a sharp interface, with no mixing of air and water, $H$ is 
a step function.  In practice, a mollified step function is used to 
provide a smooth transition between air and water.

Based on \cite{BraKotZem92, ChaHouMerOsh95}, the effects of surface 
tension are expressed as a singular source term in the Navier-Stokes 
equations:

\begin{eqnarray}
\label{tens}
T_i = \kappa(\phi) \frac{\partial}{\partial x_i} H(\phi)
\end{eqnarray}

\noindent where $\kappa$ is the curvature of the air-water interface 
expressed in terms of the level-set function:

\begin{eqnarray}
\kappa(\phi) =
\nabla \cdot \left( \frac{\nabla \phi}{| \nabla \phi |} \right) \;\; .
\end{eqnarray}

The pressure is reformulated to absorb the hydrostatic term:

\begin{eqnarray}
P & = & P_d+P_h \;\; ,
\end{eqnarray}

\noindent where $P_d$ is the dynamic pressure and $P_h$ is a 
hydrostatic pressure term:

\begin{eqnarray}
\label{static}
P_h & = & -\int^z dz' \rho(z') \frac{1}{F_r^2} \;\;
\end{eqnarray}

As discussed in \cite{DomInnLutNovSchTal98}, the divergence of the 
momentum equations (\ref{navi}) in combination with the conservation 
of mass (\ref{mass}) provides a Poisson equation for the dynamic 
pressure:

\begin{eqnarray}
\label{pois}
\frac{\partial}{\partial x_i} \frac{1}{\rho}
\frac{\partial P_d}{\partial x_i} = \Sigma \;\; ,
\end{eqnarray}

\noindent where $\Sigma$ is a source term.  Equation \ref{pois} is 
used to project the velocity onto a solenoidal field.

\section{Enforcement of Body Boundary Conditions}

Two different cartesian-grid methods are used to simulate the flow 
around the DDG 5415. The first technique imposes the no-flux boundary 
condition on the body using a finite-volume technique. The second 
technique imposes the no-flux boundary condition via an external 
force field.  Both techniques use a signed distance function $\psi$ 
to represent the body.  $\psi$ is positive outside the body and 
negative inside the body.   The magnitude of $\psi$ is the minimal 
distance between the position of $\psi$ and the surface of the body.

With respect to the volume of fluid that is enclosed by the body 
($V_b$), we define a function ${\cal F}$:

\begin{eqnarray}
{\cal F}({\bf x}) = \left\{ \begin{array}{l}
                     1 \;\; {\rm for} \;\; {\bf x} \in V_b  \\
           \frac{1}{2} \;\; {\rm for} \;\; {\bf x} \in S_b  \\
                     0 \;\; {\rm for} \;\; {\bf x} \not\in V_b \end{array}
              \right.  \;\; .
\end{eqnarray}

The function ${\cal F}$ can be expressed in terms of a surface 
distribution of normal dipoles \cite{Lam32}.

\begin{eqnarray}
{\cal F}({\bf x}) & = & \frac{1}{4 \pi} \int_{S_b} ds'
\frac{\partial}{\partial n'} \frac{1}{R} \;\; ,
\end{eqnarray}

\noindent where $n$ is the outward-pointing unit normal to the body, 
and $R$ is a Rankine source, $R  = | {\bf x} - {\bf x}' |$.  $\psi$ 
is expressed in terms of ${\cal F}$ as follows,

\begin{eqnarray}
\psi({\bf x}) =  {\cal F}({\bf x}) |{\bf x} - {\bf x}'|_{min} \;\; ,
\end{eqnarray}

\noindent where $|{\bf x} - {\bf x}'|_{min}$ is the minimal distance 
between the field point ${\bf x}$ and the points on the body ${\bf 
x}'$.  In practice, the body is discretized using triangular panels. 
As a result, the calculation of the minimal distance sweeps over all 
the triangles comprising the body and must account for the 
possibility that the minimal distance may occur either at the corners 
of triangle, along the edges of triangle, or inside the triangle.

\subsection{\label{sec:body1}Free-slip conditions}

In the finite volume approach, the irregular boundary (i.e. ship 
hull) is represented in terms of $\psi$ along with the corresponding 
area fractions $A$ and volume fractions $V$.   $V=1$ for 
computational elements fully outside the body and $V=0$ for 
computational elements fully inside the body. The representation of 
irregular boundaries via area fractions and volume fractions has been 
used previously in the following work for incompressible flows
\cite{Almgren-Bell-Colella-Marthaler:1994,UdaKanShyTay97,ColGraModPucSus99}.

Recall the pressure equation,
\be
\nabla \cdot \frac{\nabla p}{\rho}=\nabla\cdot\bmW. \label{presseqn}
\ee
with the following no-flow boundary condition:
\be
\frac{\nabla p}{\rho}\cdot \bmn_{wall}=\bmW \cdot \bmn_{wall}.
\label{noflow}
\ee
where $\bmn_{wall}$ is the outward normal drawn from the active flow 
region into the geometry region.

For each discrete computational element $\Omega_{i,j,k}$ we define 
the geometry volume fraction $V$ and area fraction $A$ as
\benl
V_{ijk} \equiv \frac{1}{|\Omega_{ijk}|}\int_{\Omega_{ijk}} H(\psi) d\bmx.
\eenl
\benl
A_{i+1/2,j,k} \equiv
   \frac{1}{|\Gamma_{i+1/2,j,k}|}\int_{\Gamma_{i+1/2,j,k}} H(\psi) d\bmx.
\eenl
$\Gamma_{i+1/2,j,k}$ represents the left face of a computational 
element; similar definitions apply to $\Gamma_{i-1/2,j,k}$, 
$\Gamma_{i,j+1/2,k}$,
\ldots.

In order to discretely enforce the boundary conditions (\ref{noflow}) 
at the geometry surface, we use a finite volume approach for 
discretizing (\ref{presseqn}).

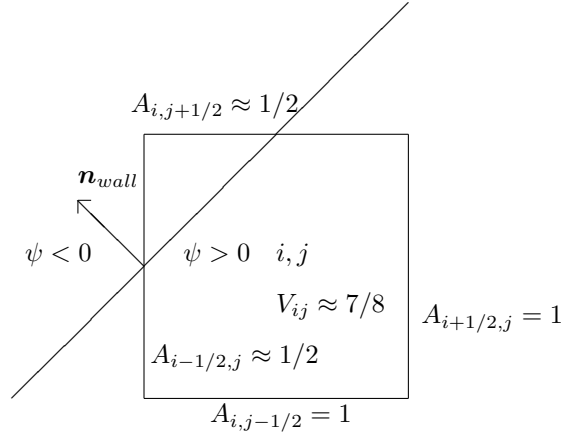
\begin{figure}[tbph]
\setlength{\unitlength}{0.5pt}

\begin{picture}(400,400)
\put(100,100){\line(1,0){200} }
\put(100,100){\line(0,1){200} }
\put(300,100){\line(0,1){200} }
\put(100,300){\line(1,0){200} }
\put(0,100){\line(1,1){300} }

\put(100,200){\line(-1,1){50} }
\put(50,250){\line(0,-1){10} }
\put(50,250){\line(1,0){10} }
\put(50,260){\makebox(0,0)[bl]{$\bmn_{wall}$} }

\put(10,200){\makebox(0,0)[bl]{$\psi<0$} }
\put(130,200){\makebox(0,0)[bl]{$\psi>0$} }
\put(200,200){\makebox(0,0)[bl]{$i,j$} }

\put(200,160){\makebox(0,0)[bl]{$V_{ij}\approx 7/8$} }
\put(150,75){\makebox(0,0)[bl]{$A_{i,j-1/2}=1$} }
\put(90,310){\makebox(0,0)[bl]{$A_{i,j+1/2}\approx 1/2$} }
\put(310,150){\makebox(0,0)[bl]{$A_{i+1/2,j}=1$} }
\put(105,120){\makebox(0,0)[bl]{$A_{i-1/2,j}\approx 1/2$} }
\end{picture}

\caption{ Diagram of computational element $(i,j)$ that is cut by the
    embedded geometry.
   \label{cutcell} }
\end{figure}

Given an irregular computational element $\Omega_{ijk}$ (see Figure 
\ref{cutcell}), we have
\benl
\int_{\Omega_{ijk}} \nabla \cdot \bmU dV =
   \int_{\partial\Omega_{ijk}} \bmU\cdot \bmn_{wall} dA.
\eenl
The divergence theorem motivates the following second order 
approximation of the divergence $\nabla \cdot \bmU$ at the centroid 
of $\Omega_{ijk}$:
\be
\nabla\cdot\bmU\approx \frac{1}{|\Omega_{ijk}|}
    \int_{\partial\Omega_{ijk}} \bmU\cdot \bmn_{wall} dA.
   \label{divthm}
\ee
In terms of geometry volume fractions $V_{ijk}$ and area fractions 
$A_{i+1/2,j,k}$, (\ref{divthm}) becomes,
\be
\lefteqn{ \nabla\cdot\bmU \approx \frac{1}{V_{ijk}\dx\dy\dz}[ }
\nonumber \\
&& (A_{i+1/2,j,k}\dy\dz) u_{i+1/2,j,k} - \nonumber \\
&& (A_{i-1/2,j,k}\dy\dz) u_{i-1/2,j,k} + \nonumber \\
&& (A_{i,j+1/2,k}\dx\dz) v_{i,j+1/2,k} - \nonumber \\
&& (A_{i,j-1/2,k}\dx\dz) v_{i,j-1/2,k} + \nonumber \\
&& (A_{i,j,k+1/2}\dx\dy) w_{i,j,k+1/2} - \nonumber \\
&& (A_{i,j,k-1/2}\dx\dy) w_{i,j,k-1/2} - \nonumber \\
&&   L^{wall}_{ijk}\bmU^{wall}_{ijk}\cdot\bmn_{wall} ] .
   \label{divthm2}
\ee
For a zero flux boundary condition at the wall, the last term in 
(\ref{divthm2}), $L^{wall}_{ij}\bmU^{wall}_{ij}\cdot\bmn_{wall}$, is 
zero.

The finite volume approach, when applied to the divergence operator 
in (\ref{presseqn}) becomes:
\benl
\lefteqn{ \nabla\cdot \frac{1}{\rho}\nabla p \approx
\frac{1}{V_{ijk}\dx\dy\dz}[ } \\
&&   A_{i+1/2,j,k}\dy\dz (p_{x}/\rho)_{i+1/2,j,k} - \\
&&   A_{i-1/2,j,k}\dy\dz (p_{x}/\rho)_{i-1/2,j,k} + \\
&&   A_{i,j+1/2,k}\dx\dz (p_{y}/\rho)_{i,j+1/2,k} - \\
&&   A_{i,j-1/2,k}\dx\dz (p_{y}/\rho)_{i,j-1/2,k} + \\
&&   A_{i,j,k+1/2}\dx\dy (p_{z}/\rho)_{i,j,k+1/2} - \\
&&   A_{i,j,k-1/2}\dx\dy (p_{z}/\rho)_{i,j,k-1/2} - \\
&&   L^{wall}_{ijk}(\nabla p/\rho)^{wall}_{ijk}\cdot\bmn_{wall}].
\eenl
and
\benl
\lefteqn{ \nabla\cdot \bmW \approx
\frac{1}{V_{ijk}\dx\dy\dz}[ } \\
&&   (A_{i+1/2,j,k}\dy\dz) u_{i+1/2,j,k} - \\
&&   (A_{i-1/2,j,k}\dy\dz) u_{i-1/2,j,k} + \\
&&   (A_{i,j+1/2,k}\dx\dz) v_{i,j+1/2,k} - \\
&&   (A_{i,j-1/2,k}\dx\dz) v_{i,j-1/2,k} + \\
&&   (A_{i,j,k+1/2}\dx\dy) w_{i,j,k+1/2} - \\
&&   (A_{i,j,k-1/2}\dx\dy) w_{i,j,k-1/2} - \\
&&   L^{wall}_{ijk}\bmW^{wall}_{ijk}\cdot\bmn_{wall}].
\eenl
Due to the no flow condition (\ref{noflow}), the terms 
$L^{wall}_{ijk}(\nabla p/\rho)^{wall}_{ijk}\cdot\bmn_{wall}$ and 
$L^{wall}_{ijk}\bmW^{wall}_{ijk}\cdot\bmn_{wall}$ cancel each other. 
The resulting discretization for $p$ is:
\benl
A_{i+1/2,j,k}\dy\dz (p_{x}/\rho)_{i+1/2,j,k} - \\
A_{i-1/2,j,k}\dy\dz (p_{x}/\rho)_{i-1/2,j,k} + \\
A_{i,j+1/2,k}\dx\dz (p_{y}/\rho)_{i,j+1/2,k} - \\
A_{i,j-1/2,k}\dx\dz (p_{y}/\rho)_{i,j-1/2,k} + \\
A_{i,j,k+1/2}\dx\dy (p_{z}/\rho)_{i,j,k+1/2} - \\
A_{i,j,k-1/2}\dx\dy (p_{z}/\rho)_{i,j,k-1/2} =
\eenl
\benl
(A_{i+1/2,j,k}\dy\dz) u_{i+1/2,j,k} - \\
(A_{i-1/2,j,k}\dy\dz) u_{i-1/2,j,k} + \\
(A_{i,j+1/2,k}\dx\dz) v_{i,j+1/2,k} - \\
(A_{i,j-1/2,k}\dx\dz) v_{i,j-1/2,k} + \\
(A_{i,j,k+1/2}\dx\dy) w_{i,j,k+1/2} - \\
(A_{i,j,k-1/2}\dx\dy) w_{i,j,k-1/2}
\eenl
where, for example,  $(p_{x})_{i+1/2,j,k}$ is discretized as
\benl
\frac{p_{i+1,j,k}-p_{i,j,k}}{\dx}.
\eenl

\subsection{\label{sec:body2}No-slip conditions}

The boundary condition on the body can also be imposed using an 
external force field.  Based on Dommermuth, et al., (1998), the 
distance function representation of the body ($\psi$) is used to 
construct a body force as follows:

\begin{eqnarray}
\label{body}
F_i({\bf x},t) = -c_f {\cal A}(t) \left( 1 -
\exp\left( -\left( \psi({\bf x})/\Delta \right)^2 \right)\right) 
u_i({\bf x},t) \nonumber \\ \forall \;\; \psi({\bf x}) \leq 0 \;\; , 
\;\;\;\;\;
\end{eqnarray}

\noindent where $c_f$ is a friction coefficient.  $\Delta$ is used to 
mollify the body force such that it is gradually applied across the 
surface of the body.  Recall that $\psi({\bf x}) \leq 0$ corresponds 
to points within the body. $F_i=0$ outside of the body.  ${\cal 
A}(t)$ is an adjustment function:

\begin{eqnarray}
{\cal A}(t) = 1.0-\exp(-(t/T_o)) \;\; .
\end{eqnarray}

\noindent $T_o$ is the adjustment time.  The adjustment function 
smoothly increases to unity from its initial value of zero.  The 
effect of the adjustment function is described in 
\cite{DomInnLutNovSchTal98}. The adjustment function reduces the 
generation of non-physical high-frequency waves.

As constructed, the velocities of the points within the body are 
forced to zero.   For a body that is fixed in a free stream, this 
corresponds to imposing no-slip boundary conditions.

\section{Interface Tracking}

Two methods are presented in our work for computing ship flows. Both 
methods use a ``front-capturing'' type procedure for representing the 
free surface separating the air and water. The first technique is 
based on the Coupled volume-of-fluid and level set method (CLS) and 
the second technique is based on the level set method (LS) alone.

\subsection{CLS method}

In this section, we describe the 2d coupled Level Set and Volume of 
Fluid (CLS) algorithm for representing the free surface. For more 
details, e.g. axisymmetric and 3d implementations, see 
\cite{SusPuc99}. In the CLS algorithm, the position of the interface 
is updated through the level set equation (level set function denoted 
by $\phi_{ij}$) and volume of fluid equation (volume fraction of 
liquid within each cell is denoted by $F_{ij}$),
\benl
  \phi_{t}+\nabla\cdot(\uvec^{MAC}\phi)=0 \\
  F_{t}+\nabla\cdot(\uvec^{MAC} F)=0.
\eenl
In order to implement the CLS algorithm, we are given a discretely 
divergence free velocity field $\bmu^{MAC}$ defined on the cell faces 
(MAC grid),
\be
\frac{u_{i+1/2,j}-u_{i-1/2,j}}{\Delta x}+
\frac{v_{i,j+1/2}-v_{i,j-1/2}}{\Delta y}=0. \label{macdiv}
\ee

Given $\phi_{ij}^{n}$, $F_{ij}^{n}$ and $\uvec^{MAC}$, we use a 
``coupled'' second order conservative operator split advection scheme 
in order to find $\phi_{ij}^{n+1}$ and $F_{ij}^{n+1}$.  The 2d 
operator split algorithm for a general scalar $s$ follows as
\be
\tilde{s}_{ij}&=&\frac{ s^{n}_{ij}+
   \frac{\dt}{\Delta x}(G_{i-1/2,j}-G_{i+1/2,j}) }
  {1-\frac{\dt}{\Delta x}(u_{i+1/2,j}-u_{i-1/2,j})}
   \label{split1}
\ee
\be
\lefteqn{s_{ij}^{n+1}=\tilde{s}_{ij}+
       \frac{\dt}{\Delta y}(\tilde{G}_{i,j-1/2}-\tilde{G}_{i,j+1/2})+ }
     \nonumber \\
& & \tilde{s}_{ij}(v_{i,j+1/2}-v_{i,j-1/2}),
   \label{split2}
\ee
where $G_{i+1/2,j}=s_{i+1/2,j}u_{i+1/2,j}$ denotes the flux of $s$ 
across the right edge of the $(i,j)$th cell and 
$\tilde{G}_{i,j+1/2}=\tilde{s}_{i,j+1/2}v_{i,j+1/2}$ denotes the flux 
across the top edge of the $(i,j)$th cell. The operations 
(\ref{split1}) and (\ref{split2}) represent the case when one has the 
``x-sweep'' followed by the ``y-sweep''.  After every time step the 
order is reversed; ``y-sweep'' (done implicitly) followed by the 
``x-sweep'' (done explicitly).
\par
The scalar flux $s_{i+1/2,j}$ is computed differently depending on 
whether $s$ represents the level set function $\phi$ or the volume 
fraction $F$.
\par
For the case when $s$ represents the level set function $\phi$ we 
have the following representation for $s_{i+1/2,j}$ ($u_{i+1/2,j}>0$),
\benl
s_{i+1/2,j}&=&s_{ij}^{n}+\frac{\Delta x}{2}(D_{x}s)_{ij}^{n}+ \\
& &  \frac{\Delta t}{2}(-u_{i+1/2,j}(D_{x}s)_{ij}^{n})
\eenl
where
\benl
(D_{x}s)_{ij}^{n} \equiv \frac{s_{i+1,j}^{n}-s_{i-1,j}^{n}}{\Delta x}.
\eenl
The above discretization is motivated by the second order predictor 
corrector method described in \cite{BelColGla89} and the references 
therein.
\par
For the case when $s$ represents the volume fraction $F$ we have the 
following representation for $s_{i+1/2,j}$ ($u_{i+1/2,j}>0$),
\be
s_{i+1/2,j}=
  \frac{  \int_{\Omega} H(\phi^{n,R}_{ij}(x,y)) d\Omega }
       {u_{i+1/2,j}\dt\dy}  \label{reconstruct}
\ee
where
\benl
\lefteqn{\Omega\equiv \{ (x,y) | x_{i+1/2}-u_{i+1/2,j}\dt\le x \le
   x_{i+1/2} } \\
& & \mbox{and} \hspace{5pt} y_{j-1/2}\le y \le y_{j+1/2} \}
\eenl
The integral in (\ref{reconstruct}) is evaluated by finding the 
volume cut out of the region of integration by the line represented 
by the zero level set of $\phi^{n,R}_{ij}$.

The term $\phi^{n,R}_{ij}(x,y)$ found in (\ref{reconstruct}) 
represents the linear reconstruction of the interface in cell 
$(i,j)$. In other words, $\phi^{n,R}_{ij}(x,y)$ has the form
\be
\phi^{n,R}_{ij}(x,y)=a_{ij}(x-x_{i})+b_{ij}(y-y_{j})+c_{ij}.
   \label{linearreconstruct}
\ee
A simple choice for the coefficients $a_{ij}$ and $b_{ij}$ is as follows,
\be
   a_{ij}=\frac{1}{2\dx}(\phi_{i+1,j}-\phi_{i-1,j})  \\
   b_{ij}=\frac{1}{2\dy}(\phi_{i,j+1}-\phi_{i,j-1}).
\ee
The intercept $c_{ij}$ is determined so that the line represented by 
the zero level set of (\ref{linearreconstruct}) cuts out the same 
volume in cell $(i,j)$ as specified by $F_{ij}^{n}$.  In other words, 
the following equation is solved for $c_{ij}$,
\benl
\frac{\int_{\Omega} H( a_{ij}(x-x_{i})+b_{ij}(y-y_{j})+c_{ij} ) d\Omega }
  { \dx\dy } = F_{ij}^{n}
\eenl
where
\benl
\lefteqn{ \Omega\equiv \{ (x,y) | x_{i-1/2}\le x \le x_{i+1/2} } \\
& & \mbox{and} \hspace{5pt}  y_{j-1/2}\le y \le y_{j+1/2} \}.
\eenl

After $\phi^{n+1}$ and $F^{n+1}$ have been updated according to 
(\ref{split1}) and (\ref{split2}) we ``couple'' the level set 
function to the volume fractions as a part of the level set 
reinitialization step. The level set reinitialization step replaces 
the current value of $\phi^{n+1}$ with the exact distance to the VOF 
reconstructed interface. At the same time, the VOF reconstructed 
interface uses the current value of $\phi^{n+1}$ to determine the 
slopes of the piecewise linear reconstructed interface.

Remarks:
\begin{itemize}

\item The distance is only needed in a tube of $K$ cells wide 
$K=\epsilon/\dx+2$, therefore, we can use ``brute force'' techniques 
for finding the exact distance.  See \cite{SusPuc99} for details.

\item During the reinitialization step we truncate the volume 
fractions to be 0 or 1 if $|\phi|>\Delta x$. Although we truncate the 
volume fractions, we still observe that mass is conserved to within a 
fraction of a percent for our test problems.

\end{itemize}

\subsubsection{CLS Contact angle boundary conditions in general geometries }

The CLS contact angle boundary conditions are enforced by extending 
$\phi$ into regions where $V_{ij}<1$ (i.e. initializing ``ghost'' 
values of $\phi$ in the inactive portion of the computational domain).

The contact angle boundary condition at solid walls is given by
\be
\bmn \cdot \bmn_{wall}=
   \cos(\theta),
\label{contactangle}
\ee
where $\theta$ is a user defined contact angle and $\bmn_{wall}$ is 
the outward normal drawn from the active flow region into the 
geometry region.

\begin{figure}[tbph]
\setlength{\unitlength}{0.5pt}

\begin{picture}(400,200)
\put(150,0){\line(0,1){200}}
\multiput(150,100)(-30,20){3}{\line(-3,2){20}}
\put(150,100){\line(3,-2){125}}
\put(150,100){\circle*{10}}
\put(90,160){\makebox(0,0)[bl]{Solid} }
\put(175,0){\makebox(0,0)[bl]{Liquid} }
\put(200,100){\makebox(0,0)[bl]{Gas} }
\put(158,70){\makebox(0,0)[bl]{$\theta$} }
\end{picture}

\caption{ Diagram of gas/liquid interface meeting at the solid.
    The dashed line
    represents the imaginary interface created thru the level-set extension
    procedure.
   \label{contactdist} }
\end{figure}
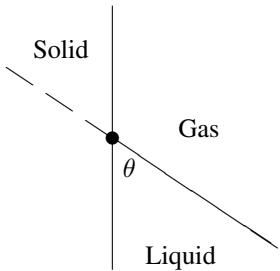

In terms of $\phi$ (the free surface level set function) and $\psi$ 
(the geometry level set function), (\ref{contactangle}) becomes
\benl
\frac{\nabla\phi}{|\nabla\phi|}\cdot \frac{-\nabla\psi}{|\nabla\psi|}=
   \cos(\theta)
\eenl
%
%
%     water | air
%           |                     nf=grad phi  nw=-grad psi
%      theta|                     nf dot nw = cos(theta)
%           |
%     nf<---|
%    ---------------------
%        solid   |
%                |
%               \ /  nw
%
%
In figure \ref{contactdist}, we show a diagram of how the contact 
angle $\theta$ is defined in terms of how the free surface intersects 
the geometry surface.

The ``extension'' equation has the form of an advection equation:
\be
\phi_{\tau}+\bmu^{extend}\cdot\nabla\phi=0 \hspace{0.5in} \mbox{$\psi<0$}
   \label{levelextend}
\ee
In regions where $\psi\ge 0$, $\phi$ is left unchanged.

For a 90 degree contact angle (the default for our computations), we have
\benl
\bmu^{extend}=-\frac{\nabla\psi}{|\nabla\psi|}.
\eenl
In other words, information propagates normal to the geometry surface.

For contact angles different from 90 degrees, the following procedure 
is taken to find $\bmu^{extend}$:
\benl
\bmn &\equiv& \frac{\nabla\phi}{|\nabla\phi|}  \\
\bmn_{wall} &\equiv& -\frac{\nabla\psi}{|\nabla\psi|} \\
\bmn_{1} &\equiv& -\frac{\bmn \times \bmn_{wall}}
                        {|\bmn \times \bmn_{wall}|} \\
\bmn_{2} &\equiv& -\frac{\bmn_{1} \times \bmn_{wall}}
                        {|\bmn_{1} \times \bmn_{wall}|} \\
c &\equiv& \bmn\cdot\bmn_{2}
\eenl
\benl
\bmu^{extend} = \left\{ \begin{array}{cc}
       \frac{\bmn_{wall}-\cot(\pi-\theta)\bmn_{2}}
           {|\bmn_{wall}-\cot(\pi-\theta)\bmn_{2}|} & \mbox{if $c<0$} \\
       \frac{\bmn_{wall}+\cot(\pi-\theta)\bmn_{2}}
           {|\bmn_{wall}+\cot(\pi-\theta)\bmn_{2}|} & \mbox{if $c>0$} \\
       \bmn_{wall} & \mbox{if $c=0$}
       \end{array}  \right.
\eenl
Remarks:
\begin{itemize}

\item In 3d, the contact line (CL) is the 2d curve which represents 
the intersection of the free surface with the geometry surface (ship 
hull). The vector $\bmn_{2}$ is orthogonal to the contact line (CL) 
and lies in the tangent plane of the geometry surface.

\item Since both $\phi$ and $\psi$ are defined within a narrow band 
of the zero level set of $\phi$, we can also define $\bmu^{extend}$ 
within a narrow band of the free surface.

\item We use a first order upwind procedure for solving 
(\ref{levelextend}). The direction of upwinding is determined from 
the extension velocity $\bmu^{extend}$. We solve (\ref{levelextend}) 
for $\tau=0\ldots \epsilon$.

\item For viscous flows, there is a conflict between the no-slip 
condition and the idea of a moving contact line.  See 
\cite{Cox86,DomInnLutNovSchTal98,HocRiv82,NgaDus89} and the 
references therein for a discussion of this issue.  We have performed 
numerical studies for axisymmetric oil spreading in water under ice 
\cite{SusUto98} with good agreement with experiments.  In the future, 
we wish to experiment with appropriate slip-boundary conditions near 
the contact line.

\end{itemize}

\subsection{Level-set method}

A key part of level-set methods is reinitialization.  Without 
reinitialization, the thickness of the interface between the gas and 
the liquid can get either too thick or too thin.   Reinitialization 
is based on the construction of a signed distance function that 
represents the distance of points from the gas-liquid interface.  By 
definition, the signed distance is positive in the liquid and 
negative in the gas.  At the interface, the distance function is 
zero.  A variety of methods have been utilized for calculating the 
signed distance function, including a hyperbolic equation 
\cite{SusSmeOsh94} and direct methods \cite{SusPuc99}. The hyperbolic 
equation methods tend to be less accurate but more efficient than 
direct methods.  Here, we outline a direct method that can be 
efficiently implemented on parallel computers with second-order 
accuracy.  The numerical scheme can also be generalized to higher 
order.

First, calculate the intersection points (${\bf x}_p$) where the zero 
level-set crosses each of the cartesian axes.  At these intersection 
points calculate the normal to the interface (${\bf n}_p$). 
Together, ${\bf x}_p$ and ${\bf n}_p$ determine local approximations 
to the planes that pass through the zero level-set.  For points that 
are within a narrow band of these planes, calculate the minimal 
distance to the planes.  Once the minimal distance is calculated, 
assign the sign of the distance function based on the sign of the 
level-set function.

For example, consider a zero crossing along the $z-$axis.   Locally, 
near the zero crossing, the level-set function $\phi$ is fitted with 
Lagrange polynomials.

\begin{eqnarray}
\tilde{\phi}(z_o) = \sum_{k=1}^{k=K} L_k(z_o) \phi_k \;\; ,
\end{eqnarray}

\noindent where $z_o$ is offset where interpolated level-set function 
$\tilde{\phi}=0$.  $L_k$ are Lagrange polynomials and $\phi_k$ are 
discrete values of $\phi$ near the zero level-set along the $z-$axis.
$K-1$ is the degree of the interpolating polynomial.   $z_o$ is 
calculated directly for low-order polynomials and iteratively for 
high-order polynomials.   Let ${\bf x_o}= (x_o,y_o,z_o)$, where 
$(x_o,y_o,z_o)$ is the coordinate of the zero crossing.

The unit normal $\bf n_o$ at the zero crossing is calculated in terms 
of the level-set function:

\begin{eqnarray}
{\bf n_o} = \frac{\nabla \phi}{|\nabla \phi |} \;\; {\rm at} \;\; 
{\bf x} = {\bf x_o} \;\; ,
\end{eqnarray}

\noindent where the gradient terms are calculated using finite 
difference formulas of desired order.

The minimal distance ($s$) between a point ($\bf x_p$) and a plane 
lies along the unit normal to the plane.  Denote the position where 
the point intersection occurs as $\bf x_s$, then

\begin{eqnarray}
{\bf x_s} = {\bf x_p} + s {\bf n_o} \;\; ,
\end{eqnarray}

\noindent where $s$ is expressed in terms of a dot product:

\begin{eqnarray}
s = ({\bf x_o} - {\bf x_p}) \cdot {\bf n_o} \;\; .
\end{eqnarray}

\noindent Note that higher-order corrections involve curvature terms, 
etc.  As long as $|{\bf x_o}-{\bf x_s}| \leq \Delta_g$, where 
$\Delta_g$ is the grid size, then $s$ is potentially the minimal 
distance to the zero level-set.  Other candidates include planes in 
the neighborhood of $\bf x_p$.  On a structured grid, shifts along the 
cartesian axes can be performed to consider other candidates.   Only 
$\bf x_p$ near the zero level-set are required in the 
reinitialization procedure.   A simple procedure for finding points 
near the zero level-set involves weighted averages.  First construct 
a stair-case approximation ($\Phi$) to the zero level-set:

\begin{eqnarray}
\Phi_{i,j,k} &  =  & 1 \;\; \forall \;\; \phi_{i,j,k} \geq 0 \nonumber \\
\Phi_{i,j,k} & = & -1 \;\; \forall \;\; \phi_{i,j,k} < 0 \;\; .
\end{eqnarray}

\noindent A weighted average along the k-th indice is

\begin{eqnarray}
\overline{\Phi}_{i,j,k}=(\Phi_{i,j,k+1}+ \Phi_{i,j,k} + 
\Phi_{i,j,k-1})/3 \;\; .
\end{eqnarray}

\noindent  Similar expressions hold along the $i-th$ and $j-th$ 
indices.    Repeated applications of weighted averages provide a 
narrow band that encompasses the zero level-set.  The narrow band 
corresponds to the region $|\overline{\Phi}_{i,j,k}| < 1$.  The 
signed distance function $D$ is expressed in terms of the level-set 
function and the minimal distance:

\begin{eqnarray}
D = {\rm sign}(\phi) \, s \;\; .
\end{eqnarray}

\noindent Based on \cite{SusSmeOsh94}, $H(\phi)$ is reinitialized as follows:

\begin{eqnarray}
H(\phi) & = & 1 \;\; {\rm if} \;\; D > \Delta \nonumber \\
H(\phi) & = & \sin( \frac{\pi D}{2 \Delta}) \;\; {\rm if} \;\;  |D| 
\leq \Delta  \nonumber \\
H(\phi) & = & -1 \;\; {\rm if} \;\; D < -\Delta \;\; ,
\end{eqnarray}

\noindent where $\Delta$ is the desired thickness of the interface.

\section{Flux Integral Methods}

We define the temporal and spatial averaging over a time step and a 
cell as follows:

\begin{eqnarray}
\widetilde{\overline{\phi}} = \frac{1}{\Delta t \Delta V} 
\int_t^{t+\Delta t} dt \int_V dv \, \phi \;\; ,
\end{eqnarray}

\noindent where here, the tilde and overbar symbols respectively 
denote temporal and spatial averaging. $\Delta t$ is the time step, 
and $\Delta V$ is the volume of the cell.

As an example, consider the application of the preceding operator to 
the level-set equation (\ref{level}):

\begin{eqnarray}
\label{fim1}
\frac{\overline{\phi}^{n+1} - \overline{\phi}^n}{\Delta t} + 
\widetilde{\overline{\frac{\partial u_j \phi}{\partial x_j}}}= 
\widetilde{\overline{\frac{\partial Q}{\partial x_j}}} \;\; ,
\end{eqnarray}

\noindent where here superscript $n$ denotes the time level.  We 
focus our attention on the convective term.  The convective term 
accounts for the flux of the level-set function across the faces of 
the control volume.  A second-order approximation for the flux across 
one face of a cell is provided below:

\begin{eqnarray}
\label{fim2}
F_x^+ = \int_{x_1}^{x_2} dx \int_{y_1}^{y_2} dy \int_{z_1}^{z_2} dz 
\,\,\, \phi(x,y,z) \;\; ,
\end{eqnarray}

\noindent where $F_x^+$ is the flux across the positive face along 
the x axis.  The limits of integration are provided below:

\begin{eqnarray}
\label{fim3}
x_1 & = & \frac{\Delta x}{2} \nonumber \\
x_2 & = & \frac{\Delta x}{2} - u^+ \Delta t \nonumber \\
y_1 & = & -\frac{\Delta y}{2} + (x-\frac{\Delta x}{2})\frac{v^-}{u^+} 
\nonumber \\
y_2 & = & \frac{\Delta y}{2} +  (x-\frac{\Delta x}{2})\frac{v^+}{u^+} 
\nonumber \\
z_1 & = & -\frac{\Delta z}{2} + (x-\frac{\Delta x}{2})\frac{w^-}{u^+} 
\nonumber \\
z_2 & = & \frac{\Delta z}{2} + (x-\frac{\Delta x}{2})\frac{w^+}{u^+} \;\; ,
\end{eqnarray}

\noindent where $\Delta x$, $\Delta y$, and $\Delta z$ are the 
lengths of the cell along the cartesian axes.  $u^+$ is the normal 
component of fluid velocity at the center of positive face along the 
$x-$axis.  $v^+$ and $v^-$ are the normal components of the fluid 
velocities at the centers of the positive and negative faces along 
the $y-$axis.  Similar definitions hold for $w^+$ and $w^-$.  

In a mapped coordinate system, the expression for the flux is

\begin{eqnarray}
\label{fim4}
F_x^+ = \int_{-1}^{1} \int_{-1}^{1} \int_{-1}^{1} dr  ds  dt \,\,\, J 
\; \phi(x,y,z) \;\; ,
\end{eqnarray}

\noindent where $J$ is the jacobian, and $x$, $y$, and $z$ are 
functions of $r$, $s$, and $t$:

\begin{eqnarray}
\label{fim5}
x & = & \frac{\Delta x}{2} - \frac{(1+r)}{2} u^+ \Delta t \nonumber \\
y & = & \frac{s \Delta y}{2} - \frac{(1+r)(1+s) v^+ \Delta t}{4} \nonumber \\
&& -\frac{(1+r)(1-s) v^- \Delta t}{4} \nonumber \\
z & = & \frac{t \Delta z}{2} - \frac{(1+r)(1+t) w^+ \Delta t}{4} \nonumber \\
&& -\frac{(1+r)(1-t) w^- \Delta t}{4} \;\; .
\end{eqnarray}

\noindent For this particular approximation, the jacobian is
\begin{eqnarray}
\label{fim6}
J & = & \frac{\partial x}{\partial r} \frac{\partial y}{\partial s}
\frac{\partial z}{\partial t} \;\; .
\end{eqnarray}

\noindent On any one face the stencil associated with the Lagrangian 
interpolation of $\phi$ is $3\times3\times3=27$ points, but for the 
entire cell, the stencil is $5\times5\times5=125$ points.  We use a 
upwind-biased stencil for the momentum equations and a symmetric 
stencil for the level-set function.  The diagonal and cross terms in 
the momentum equations are treated the same.  Generally, we use 
eight-point Gaussian quadrature to evaluate the flux over each face. 
Details of the numerical algorithm are described in \cite{Dom00}. 
Various types of limiters are described in \cite{Leo97}.

\section{Preliminary Results}

In section \ref{shipwaveresults}, we present preliminary computations 
of flow past a DDG 5415 ship.  In section \ref{sprayresults}, we 
present preliminary computations of the breakup of a two-dimensional 
spray sheet.

\subsection{Ship Wave Results \label{shipwaveresults} }

As a demonstration of the level-set and the coupled level-set and 
volume-of-fluid formulations, we predict the free-surface disturbance 
near the bow of the DDG 5415 moving with forward speed.  The 
experiments were performed at the David Taylor Model Basin (DTMB), 
and are available via the world wide web at 
http://www50.dt.navy.mil/5415/.  This is the same flow that 
Dommermuth, et al., (1998) originally investigated using their 
stratified flow formulation \cite{DomInnLutNovSchTal98}.  As before, 
we only consider the high speed case. For this case, a plunging 
breaker forms near the bow.   Air is entrained and splash up occurs 
where the wave reenters the free surface.  There is flow separation 
at the stern, and the transom is dry.  A large rooster tail forms 
just behind the stern.

Based on the speed ($U_o$=6.02Knots) and the length ($L_o$=5.72m) of 
the model, the Reynolds and Froude numbers are $R_e=1.8\times10^7$ 
and $Fr^2=0.41$.  The effects of surface tension are not included. 
The density ratio of air and water is $\lambda=0.0012$ and the ratio 
of the dynamic viscosities is $\eta=0.018$.

In regard to the numerical parameters for the level-set formulation, 
we use a friction coefficient $c_f=500$ in the body-force term 
(\ref{body}).   The adjustment time is $T_o=0.02$.  For the level-set 
formulation, the length and width of the computational domain are 
$L=2.5$ and $W=1.50$.  The height of the air above the mean 
free-surface is $h=0.15$ and the depth below the mean free-surface is 
$d=1.0$.  One grid resolution is used with $512\times128\times129$ 
grid points.  Three different levels of grid stretching are used 
along the $y-$ and $z-$axes.  For the highest grid resolution, the 
smallest grid spacing is $2.6\times10^{-3}$ along the $y-$axis and 
$3.6\times10^{-4}$ along the $z-$axis.  For the medium resolution 
simulation, the smallest grid spacing is $3.8\times10^{-3}$ along the 
$y-$axis and $1.8\times10^{-3}$ along the $z-$axis.  For the coarsest 
grid simulation, the smallest grid spacing is $3.8\times10^{-3}$ 
along the $y-$axis and $3.5\times10^{-3}$ along the $z-$axis.  The 
grid spacing ($4.9\times10^{-3}$) is constant along the $x-$axis for 
all three cases.  The thicknesses of the free-surface interfaces for 
the fine, medium, and coarse simulation are respectively $\Delta=0.05$, 
$0.025$, and $0.0125$.  The durations of the coarse and medium resolution 
simulations are $t=0.76$ and $t=0.68$, respectively.  No 
special treatment is used for the level-set function inside the ship. 
These durations correspond to about three quarters of a ship length 
based on the present normalization.  For these durations, the flow is 
steady near the bow and still evolving near the stern.  (The fine 
resolution simulation is still evolving, and it is not possible at 
this time to present complete results.  More complete results will be 
provided at the symposium and in the discussion section of this paper 
\footnote{We would have performed longer simulations, but the NAVO 
T3E was unexpectedly shutdown for five days of maintenance just 
before this paper was due.}.)  The ship is centered in the 
computational domain with the same fixed sinkage and trim as used in 
the experiments.  In order to construct the body force term, the hull 
is panelized using approximately $4000$ panels.

Coarse and medium resolution simulations have been performed using 
the CLS formulation.  The coarse simulation uses 
$256\times64\times64$ grid points, and the fine resolution uses 
$512\times128\times128$ grid points.  The length, width, and height 
of the computational domain are $L=2$, $W=0.5$, and $H=0.5$, 
respectively.  The water depth is $d=0.25$.  The grid spacing is 
constant along all three cartesian axes.  In the next phase of our 
research, we will implement grid stretching, which will allow greater 
water depths to be simulated.  The durations of the CLS simulations 
are $t=0.75$.   Unlike the level-set results, the CLS results extend 
the free-surface interface into the hull using the techniques 
outlined earlier in our paper.

The free-surface elevation was measured at DTMB using a whisker 
probe. Twenty-one transverse cuts were performed near the bow, 
extending from $x=0$ to $x=0.178$ in dimensionless units.  The 
whisker probe measures the highest point of the free surface.  In 
regions where there is wave breaking, the whisker probe measures the 
top of the breaking wave.  Seventeen transverse cuts were performed 
in the stern, extending from $x=1.01$ to $x=1.22$.

\begin{figure}[htbp]
\centerline{\includegraphics[width=0.99\linewidth]{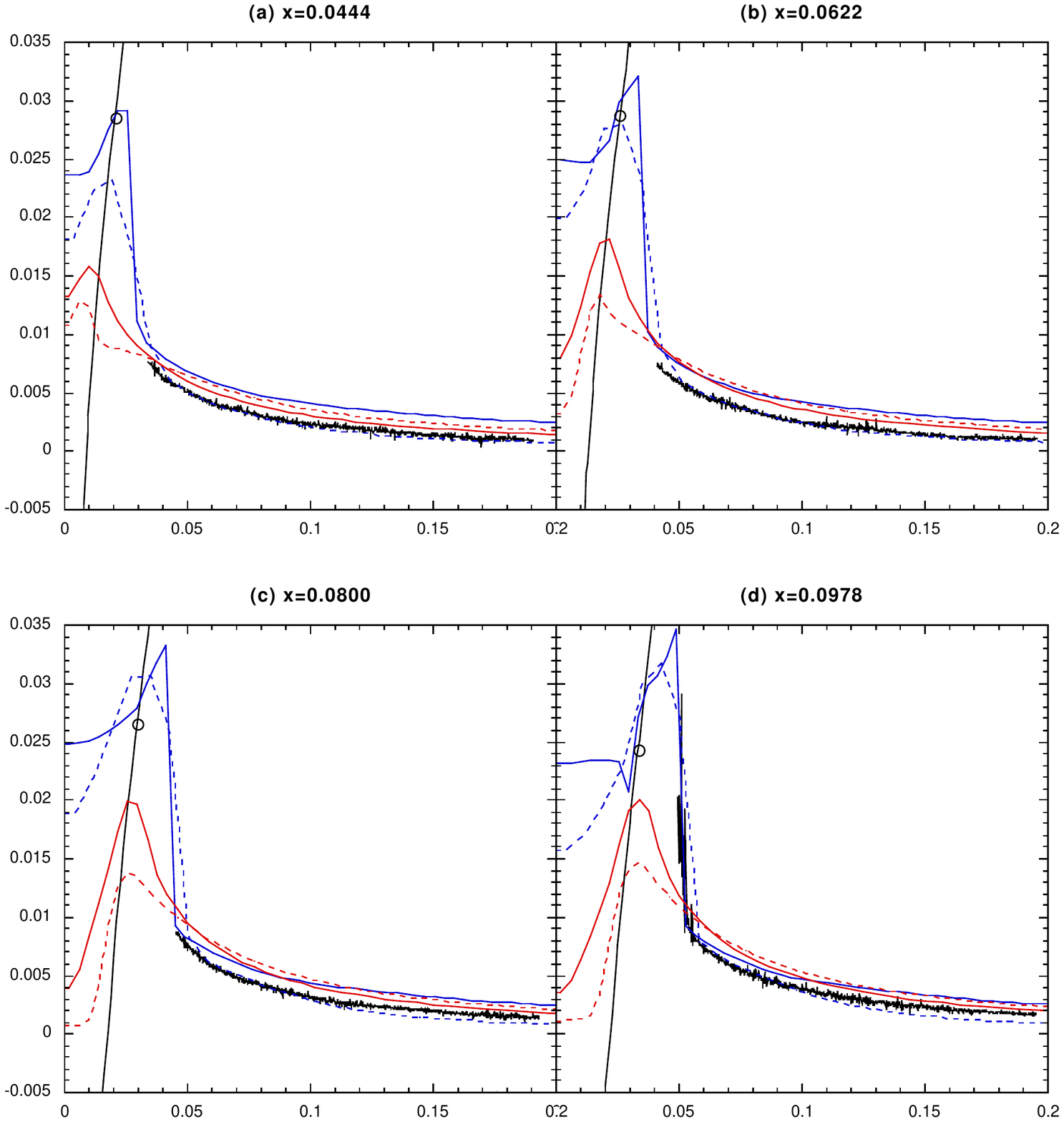}}
\caption{Flow near bow. \label{bow_fig} }
\end{figure}

\begin{figure}[htbp]
\centerline{\includegraphics[width=0.8\linewidth]{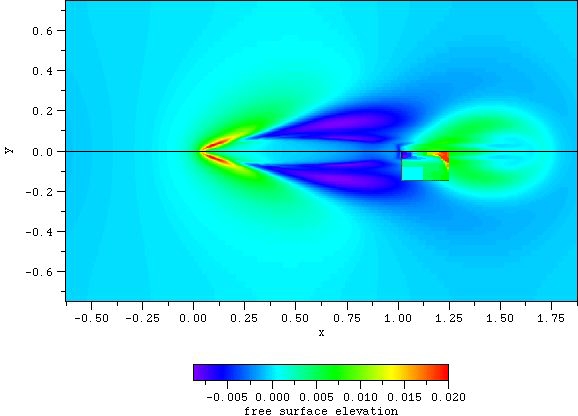}}
\caption{Flow near stern. \label{stern_fig} }
\end{figure}

Figures \ref{bow_fig} and \ref{stern_fig} compare measurements at the 
bow and stern to the numerical predictions.  The bow measurements 
include profile and whisker-probe measurements.   Comparisons to the 
bow data are performed at four stations: $x=0.0444$, $x=0.0622$, 
$x=0.0800$, and $x=0.0978$.  The circular symbol denotes profile 
measurements.  The solid black lines denote the outline of the hull 
and the whisker-probe measurements.  The solid blue line is medium 
CLS and the dashed blue line is coarse CLS.  The solid red line is 
medium level-set and the dashed red line is coarse level-set.  In 
general, the CLS technique captures the rapid rise up the side of the 
hull.  The level-set technique does less well in this regard.  In the 
outer-flow region the CLS coarse results are slightly better than the 
CLS fine results.  This may be attributed to the shallow depth that 
is used in the CLS.  The level-set results appear to converge better 
in the outer-flow region, but the results of the fine simulation are 
required for confirmation.

Figure \ref{stern_fig} shows the entire flow around the ship for the 
medium resolution level-set simulation. The stern whisker-probe 
measurements are overlaid for the purposes of comparison.  Although 
the numerical  results are not stationary, the shape of the stern 
contours show general agreement with laboratory measurements. 
However, the amplitude of the numerical results are significantly 
lower than the measurements.  Note that the stern is partially dry in 
the numerical simulations.  The outline of the hull is visible in the 
numerical simulations because the level-set function intersects the 
hull.

\subsection{Spray Sheet Results \label{sprayresults} }

The Navier-Stokes equations in combination with a level-set 
formulation are used to study the breakup of two-dimensional sheet of 
water.   The sheet is $l_o=6$mm thick.  The length of the sheet is 24mm.  
The top and bottom of the sheet are bounded by air.  The initial mean-velocity 
of the water is $u_o=3$m/s.  The initial rms turbulent 
velocity of the water is $\tilde{u}=1.2$m/s.   The air is initially 
quiescent.  Based on the sheet thickness ($l_o$) and the mean 
velocity ($u_o$), the Reynolds number is $R_e=u_o l_o / \mu=18,000$ 
and the Weber number is $W_e= \rho u_o^2 l_o/\sigma=730$,  where 
$\mu$ is the kinematic viscosity of water, $\rho$ is the water 
density, and $\sigma$ is the surface tension.  The density and 
viscosity ratios are $\lambda=\rho_g/\rho_\ell=0.0012$ and 
$\eta=\mu_g/\mu_\ell=0.018$, which are appropriate for air-water 
interfaces.  This parameter regime roughly corresponds to experiments 
that were performed by Sarpkaya and Merrill (1998), \cite{SarMer98}. 
Numerical convergence is established using $2048^2$ and $4096^2$ grid 
points.   Second-order accuracy in space is established.  A 
third-order Runge-Kutta scheme is used to integrate the system of 
equations with respect to time.  Mass is conserved to within 0.25\% 
throughout the entire calculation.

\begin{figure}[htbp]
\centerline{\includegraphics[width=0.9\linewidth]{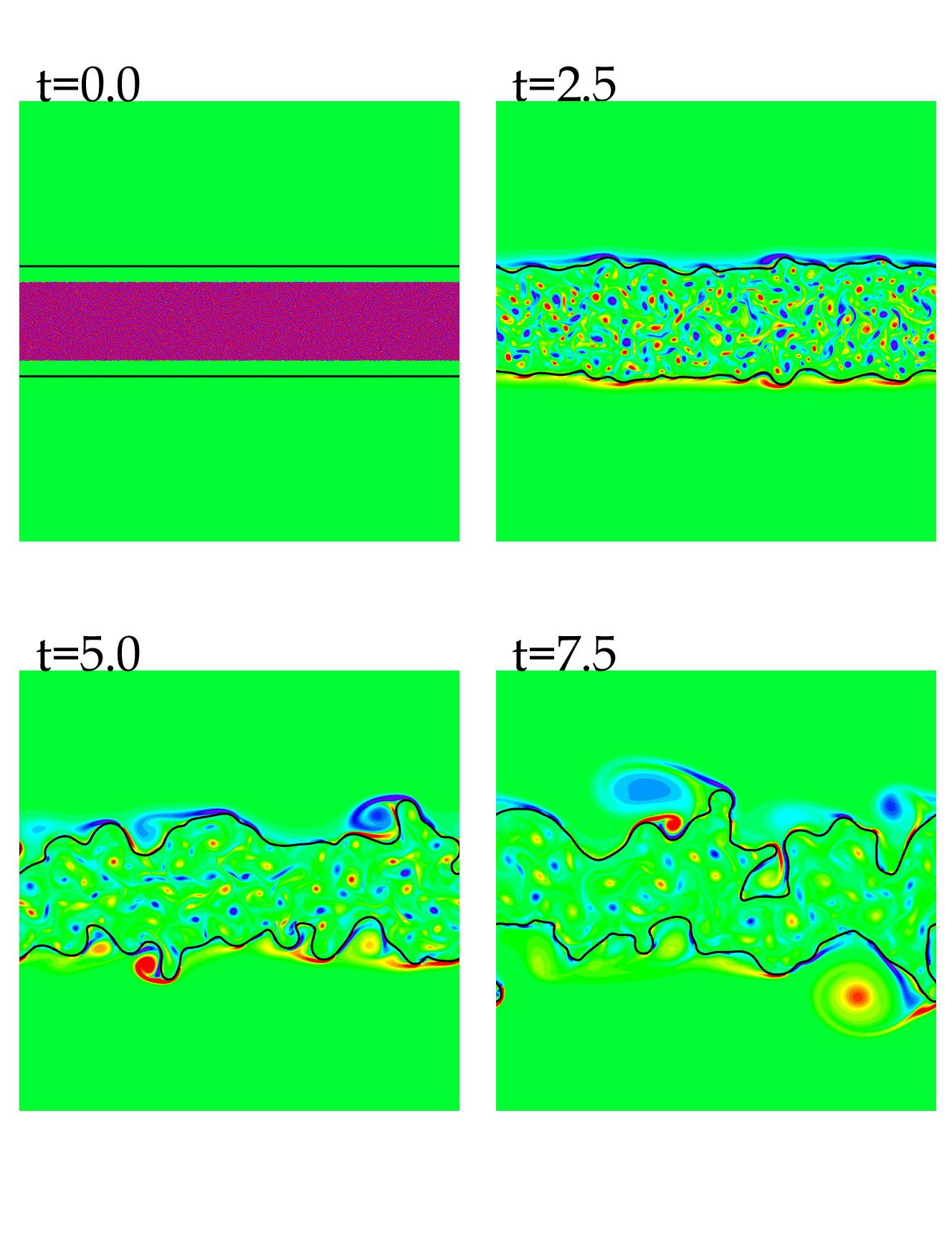}}
\caption{2d spray sheet. \label{sheet_fig} }
\end{figure}

Figure \ref{sheet_fig} illustrates the evolution of a two-dimensional 
spray sheet.  The black contour lines indicate the interface between 
air and water.   The water sheet is bounded by air both at the top 
and the bottom of the sheet. The color contours denote the vorticity. 
The flow is turbulent within the water sheet and laminar in the air. 
The mean velocity and rms velocity profiles are initially top-hat 
functions. The flow is moving from left to right. The turbulent 
fluctuations in the water are initially immersed below the top of the 
sheet and above the bottom of the sheet (see Fig.\ 
\ref{sheet_fig}:$\, t=0$).

The turbulence in the water diffuses and interacts with the 
interfaces (see Fig.\ \ref{sheet_fig}:$\, t=2.5$).  The initial 
interaction is a roughening of the air-water interface.  A thin 
boundary layer forms in the air.  The boundary layer is colored blue 
(negative) at the top of the sheet and colored red (positive) at the 
bottom of the sheet. As the interface gets rougher and ligaments 
begin to form, the air separates from the back of the ligaments.  The 
boundary layer thickens, and air is dragged along the top and the 
bottoms of the sheet.

Primary vortex shedding initially occurs behind the ligaments (see 
lower left of sheet in Fig.\ \ref{sheet_fig}:$\, t=5$).  As the 
primary vortices are shed, their interactions lead to the formation 
of secondary and tertiary vorticity (see upper middle of sheet in 
Fig.\ \ref{sheet_fig}:$\, t=7.5$).  Vortices are periodically shed 
from the backs of ligaments (see lower middle of sheet in Fig.\ 
\ref{sheet_fig}:$\, t=10$).  There is evidence of vortex merging both 
in the air and in the water (see upper left of Fig.\ 
\ref{sheet_fig}:$\, t=17.5$).  Although there is significant flow 
separation in the air, there is little or no separation in the water. 
The largest ligaments are formed by eddies impinging on the interface 
(see upper left of Fig.\ \ref{sheet_fig}:$\, t=12.5$).  Cavities form 
in regions where primary vortices are trapped.  The inlets to the 
cavities shed secondary vorticity, which tends to make the cavities 
even larger (see middle of sheet in Fig.\ \ref{sheet_fig}:$\, t=15$). 
At the inlets to the cavities, vortex pairs are formed.  Under their 
own self-induced velocities, the vortex pairs move into the cavities 
where they diffuse.

Note that droplets do not actually form at the tips of the ligaments 
because 2d flows are not subject to the same instabilities as 3d 
flows.  The turbulent kinetic energy tends to concentrate in the 
thicker portions of the deformed spray sheet.  The flow within the 
ligaments is relatively benign.   In agreement with theory, the 
pressure at the tips of the longest ligaments roughly scales like 
$P=(W_e r)^{-1}$, where $r$ is the radius of curvature of the tip.

\section{Conclusion}

In this paper, we have outlined the key numerical algorithms for 
simulating free-surface flows on cartesian grids using level-set and 
coupled level-set and volume-of-fluid techniques.  Preliminary 
numerical results have been shown for ship waves and spray sheets. 
The ship wave results indicate that cartesian-grid methods are 
capable of resolving the flow around a ship if the grid resolution is 
sufficient.  Near the bow and stern, we estimate that the grid 
spacing along all three cartesian axes should be $\Delta=0.0005$ 
(based on ship length) in order to resolve breaking waves.  On a 
parallel computer, it is possible to approach this level of grid 
resolution, but adaptive gridding may also be required to fully 
resolve the entire flow around a ship \cite{SusAlmBelColHowWel99}. 
Alternatively, cartesian-grid methods could be embedded in more 
conventional boundary-fitted methods to capture complex flows near 
the bow or stern.  The spray-sheet results show that cartesian-grid 
methods are capable of resolving the air and water boundary layer at 
realistic Reynolds numbers.

{\it Acknowledgments.}  The first author is supported in part by NSF 
Division of Mathematical Sciences under award number DMS 9996349. 
The second author is supported by ONR under contract number 
N00014-97-C-0345. Dr.\ Edwin P. Rood is the program manager. The 
numerical simulations have been performed on the T3E computer at the 
Naval Oceanographic Office using funding provided by a Department of 
Defense Challenge Project.  We are very grateful to Mr.\ George 
Innis, Dr.\ James Rottman, and Mr.\ Andrew Talcott for assistance 
with this paper.

\bibliography{references}
\bibliographystyle{plain}

\newpage
\onecolumn

\begin{figure}[htbp]
\centerline{\includegraphics[width=0.9\linewidth]{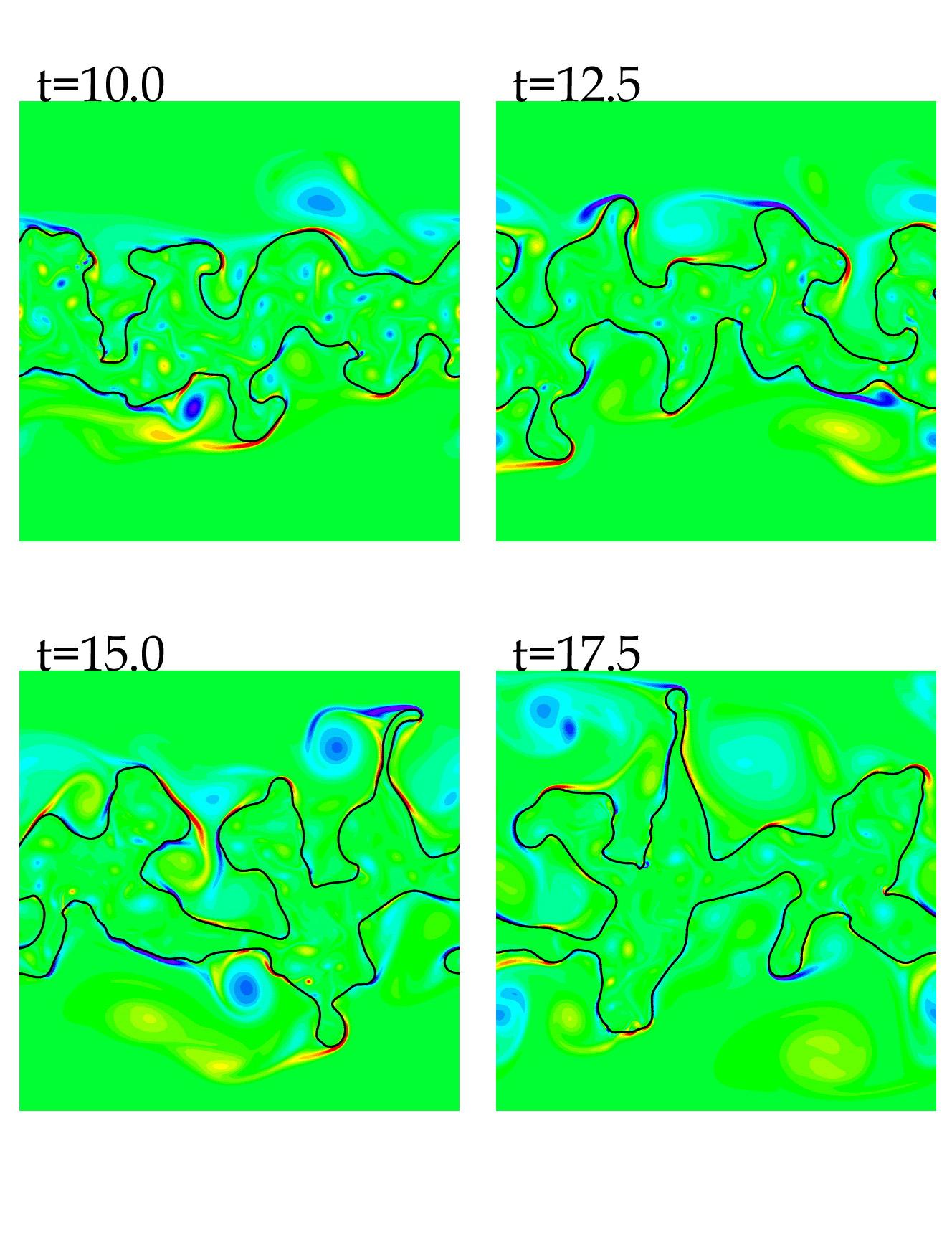}}
\begin{center}
Figure \ref{sheet_fig}: 2d spray sheet continued.
\end{center}
\end{figure}

\end{document}